\documentclass[%
reprint,
superscriptaddress,
showkeys,
nofootinbib,
amsmath,amssymb,
aps,
floatfix,
]{revtex4-2}
\usepackage{accents}
\usepackage{graphicx}
\usepackage{dcolumn}
\usepackage{bm}
\usepackage[utf8]{inputenc}
\usepackage{tabularx}
\usepackage[dvipsnames]{xcolor}
\usepackage{units}
\usepackage[euler]{textgreek}
\usepackage{upgreek}
\usepackage{derivative}
\usepackage{amsmath}
\usepackage{physics}
\usepackage{amssymb}
\usepackage{appendix}
\usepackage[colorlinks=true,linkcolor=blue!50!black,urlcolor=blue!50!black,citecolor=blue!50!black]{hyperref}

\graphicspath{ {img/} }
\usepackage{comment}

\usepackage{dsfont}

\newcommand{\uuline}[1]{{\boldsymbol{#1}}}

\def\d{\textrm{d}}

\def\eg{\textit{e.g.}}
\def\pI{{\phantom{I}}}

\usepackage{parskip}
\begin{document}


\preprint{APS}

\title{The relative interfacial thermal contraction as a possible origin\\ of the low energy excess in cryogenic calorimeters}

\author{Vanessa Zema}
\email{vanessa.zema@mpp.mpg.de}
\email{vanessa.zema@oeaw.ac.at}
\affiliation{Max Planck Institute for Physics, 85748 Garching - Germany}
\affiliation{Marietta Blau Institute for Particle Physics of the Austrian Academy of Sciences, 1010 Wien - Austria}

\author{Pasquale Pavone}
\email{pasquale.pavone@physik.hu-berlin.de}
\affiliation{Department of Physics and CSMB, Humboldt-Universität zu Berlin, 12489 Berlin - Germany}

\begin{abstract}
Low threshold cryogenic calorimeters are a key technology for the advancement of rare-event searches. However, since a few years their sensitivity reach is challenged by the presence of a rising spectrum at low energies named low-energy excess (LEE), ascribed to an unknown background.
In this work, we describe the LEE as absorber events induced by the relative thermal-contraction coefficient mismatch between the absorber and the SiO$_2$ amorphous layer underneath the transition-edge sensors (TESs), present for example in the case of CRESST detectors. The relative contraction in processes with temperature changes, such as during sensor fabrication and cooldown from room temperature to the temperature of operation, can induce surface dislocation nucleation. Other interfaced materials with thermal-expansion mismatch can also generate dislocations during temperature-variation processes. We formulate a simple elastic model to bridge this solid-state effect and the LEE observations. Double-TES modules have been designed to provide surface background rejection. We highlight that the presence of the LEE in the coincident event band of double-TES modules does not exclude the explanation given in this work. Exemplary, we discuss the role of the thermal boundary resistance between absorber and sensor as explanation for the presence of the LEE in the coincident-event band. We propose detector designs to test these hypotheses and mitigate the LEE.

\end{abstract}

\maketitle

\section{Introduction}
A cryogenic calorimeter consists of a crystal target named \textit{absorber} and a phonon sensor in thermal contact with the absorber surface. Detectors based on cryogenic calorimetry are commonly employed for rare-event searches because of the exceptional energy resolution at low energies. However, the application of cryogenic calorimeters for rare-event searches is currently challenged by the observation of an unexpected rise of the event rate at low energies, named low-energy excess (LEE)~\cite{Fuss:2022fxe, Baxter:2025odk}.
CRESST-III has excluded dark matter and any other external radiation as major sources because the observed LEE does not scale with the mass of the absorber and decays with time~\cite{Angloher:2022pas, Baxter:2025odk}. Also, by performing thermal cycles at different temperatures, EDELWEISS~\cite{EDELWEISS:2021}, CRESST-III~\cite{Angloher:2022pas} and others~\cite{Baxter:2025odk} showed that a component of the LEE rate can be recharged by warming up the cryostat, and subsequently decays with fast decay times after data taking at base temperature is resumed. NUCLEUS has recently reported on observations indicating a correlation of the initial LEE rate with the time duration of the cooldown, showing lower initial rates with slower cooldowns~\cite{Abele:2026sqm}. CRESST-III data taken right after a neutron calibration shows that radiation damage induced by neutron scattering does not have an impact on the LEE~(\cite{Angloher:2022pas}, Figure~5). Another observation is that the LEE events have the same pulse shape as absorber events~\cite{Angloher:2022pas}. TESSERACT has recently reported on two measurements dedicated to LEE investigations: The first showing a LEE which increases with increasing target volume, for energies below about 30\:eV~\cite{TESSERACT:2025odn}; the second showing that fast-neutron induced defects are not a major source of their LEE background~\cite{Armatol:2026wro}. The difference in the LEE mass scaling rule between CRESST-III and TESSERACT may point towards the presence of different components of the LEE contributing to the overall energy spectrum.

The hypothesis investigated in this work,\footnote{Part of this work was already presented in the context of the EXCESS workshop, see \eg,~\cite{Zema2023, Baxter:2025odk}} which fulfills all observations (except for the mass scaling observed by TESSERACT below 30 eV which we address below), is that part of the LEE originates from the mismatch between the thermal-expansion coefficient (TEC) of both sensor and absorber. In the specific case of CRESST, the stress may generate from the mismatch between the TEC of the amorphous SiO$_2$ layer underneath the sensor and the absorber. Such a mismatch would induce a shear stress at the interface, which above a certain threshold displacement energy can result in \textit{surface dislocation nucleation}~\cite{PhysRevLett.100.025502, ARSENAULT1986175}. 

Studies of bulk dislocation nucleation connected to the thermal expansion mismatch between crystal lattice and contaminants or defects (\textit{inclusions}) are also present in the literature~\cite{ARSENAULT1986175}.~This mechanism may contribute to the mass-scaling behavior observed by TESSERACT.

The relaxation of these dislocations may cause the observed LEE, with mechanisms similar to the lattice defect relaxations discussed in~\cite{p7s5-4qjw}. However, since~\cite{p7s5-4qjw} refers to lattice damages induced by radiation and not by thermoelastic stress, dedicated atomistic simulation based on molecular dynamics and density-functional theory are ongoing, motivated by this work.

An experimental effort towards the mitigation of the LEE via surface-background rejection is the development of double-TES readouts, consisting in target crystals equipped with two TESs or two TESs lines fabricated on their surfaces~\cite{GRAPES-3:2024yym, TESSERACT:2025odn}. Events in coincidence are interpreted as particle events; events in only one TES are rejected because they are ascribed to surface backgrounds. In this work we motivate why these experimental results do not exclude that the TEC absorber/sensor mismatch can induce LEE events, on the basis of the phenomenology of phonon propagation and thermal boundary resistance mismatch between absorbers and sensors.

In Section~\ref{sec:TEC} we present a theoretical simple model to predict the energy produced during a cryogenic cooling by the TEC mismatch and compare to the energy scale of the LEE spectrum. In Section~\ref{sec:dislocation} we describe surface dislocation nucleation and threshold dislocation energy, largely described in material science and engineering and we contextualize this research to our case of study. In Section~\ref{sec:acma} we report on the possible effect of thermal boundary resistance mismatch at the interface on the attempted surface-event-rejection with double-TES readout modules. In Section~\ref{sec:proposal} we propose tests of these hypotheses and detector designs which could mitigate or ideally eliminate the LEE.

\section{Thermal-expansion coefficient mismatch} 
\label{sec:TEC}
The linear TEC at constant pressure is defined~as~\cite{AschMerm_chap25},
\begin{equation} 
\label{eq:alpha}
    \alpha (T) = \frac{1}{L}\left(\frac{\partial L}{\partial T}\right)_{\!\!P},
\end{equation}
where $L$ is the length, $T$ the temperature, and $P$ the pressure.
\begin{figure}[t!]
\includegraphics[width=0.47\textwidth]{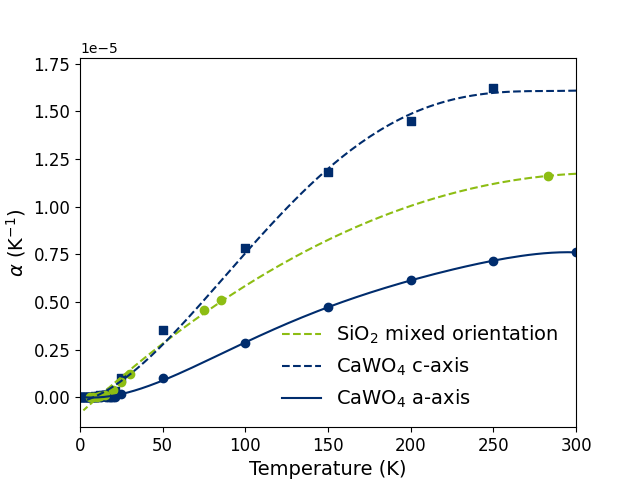}
\vspace{-2mm}
\caption{Thermal-expansion coefficient of SiO$_2$ with mixed orientation and CaWO$_4$ as a function of temperature. Additional points have been kindly provided by A.~Senyshyn, author of~\cite{senyshyn2004lattice}, under explicit request for this study. The data points are from~\cite{senyshyn2004lattice} for CaWO$_4$ and~\cite{white1964thermal} for SiO$_2$ with mixed orientation.} 
\label{fig:TECs_SiO2_CaWO4}
\end{figure}
Measurements of the linear TEC in different materials at low temperatures are available in the literature~\cite{white1964thermal,corruccini1961thermal, barron1980thermal, senyshyn2004lattice}.
In Fig.~\ref{fig:TECs_SiO2_CaWO4} we report the linear TEC of CaWO$_4$ measured along the $c$-axis (dark-blue squared markers), fitted with a polynomial function (dark-blue dashed curve) and the one along the $a$-axis (dark-blue round markers) fitted with a polynomial function (dark-blue solid line)~\cite{senyshyn2004lattice}. These data are also compared to the ones of SiO$_2$ with mixed orientation~\cite{white1964thermal} (green markers and dashed curve). Assuming that the experimental values reported in literature are comparable to the ones of the CRESST modules, Fig.\:\ref{fig:TECs_SiO2_CaWO4} shows a mismatch of the thermal contraction which could induce a shear stress at the interface between the absorber and the SiO$_2$ layer. It is evident that the TECs of CaWO$_4$ along the $c$-axis and SiO$_2$ with mixed orientation are similar below 50\:K and that the differences further decrease at lower temperatures. Focusing on two different modules of CaWO$_4$, a clear recharged LEE after a 30\:K warm-up was observed for one module but not for the other, while both resulted in a recharged LEE after a warm-up to 60\:K~\cite{Zema2023, dissertation}. One module has a commercial CaWO$_4$ crystal, labeled "comm2", and the other one a CaWO$_4$ crystal grown at the Technical University of Munich (TUM) and labeled TUM93A~\cite{dissertation}. A possible explanation is that the two modules have different orientations, \textit{e.g.}, along the $a$- and $c$-axis and that the one which does not show a recharged LEE at the 30\:K warm-up is oriented along the $c$-axis. The orientation of CRESST crystals was not measured, as it is considered irrelevant for the TES production, when the SiO$_2$ amorphous layer is fabricated in between absorber and TES - a measurement of the orientation of the crystals in these two modules could test this hypothesis. Additionally, the warm-up tests performed by CRESST-III showed in different modules (Si, Al$_2$O$_3$, and CaWO$_4$) evidence for recharged excess for warm-ups above 30\:K, but not below 11\:K~\cite{Zema2023, kuckuk2025lee}, observation which further supports the hypothesis of this work. 

\begin{figure*}[htb!]
\includegraphics[width=\linewidth]{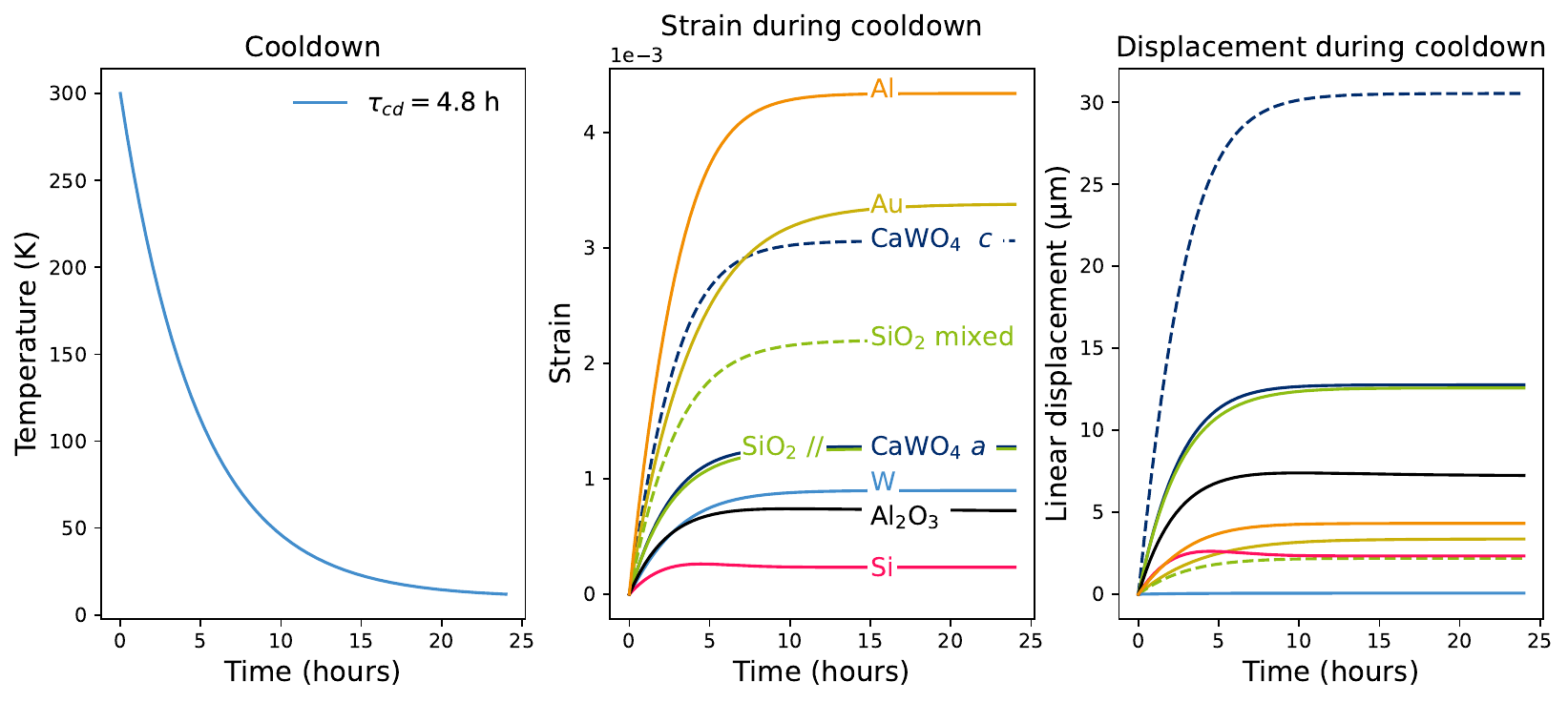}
\vspace{-6mm}
\caption{Thermal properties of different materials during a cooldown exponential in time with time constant $\tau_{cd} = 4.8$~h. Left: Cooldown temperature curve in time. Center: Strain during cooldown. L(300 K) is 1 cm for CaWO$_4$-a/c, Al$_2$O$_3$, SiO$_2$ and Si, 0.1 cm for Al, Au and SiO$_2$ mixed, and 0.007 cm for W.  Right:
Displacement during cooldown.}
\label{fig:cooldown}
\end{figure*}

To quantify the order of magnitude of the elastic energy associated to the thermal contraction of the absorber, we use the classic theory of elasticity for crystals. We start from the expression of the elastic energy density,~$u$, as a Taylor expansion in the components of the physical-strain tensor~$\uuline{\epsilon}$:
\begin{equation}
\label{eq:UoverV}
u(\uuline{\epsilon}) = u_0^\pI+\frac{1}{2} \, \sum_{ij.kl} C_{ij,kl} \, \epsilon_{ij} \, \epsilon_{kl}+\mathcal{O}(\epsilon^3)\:,
\end{equation}
where $u_0^\pI$ is the energy density at vanishing strain, the indices $i,j,k,$ and $l$ represent the Cartesian coordinates, $C_{ij\,kl}$ are the components of the fourth-rank elastic tensor $\uuline{C}$ of the absorber, and $\epsilon_{ij}$ are components of the strain tensor~\cite{AschMerm_chap25}. The crystal symmetry of the CaWO$_4$ is tetragonal~\cite{zalkin1964x, FARLEY1971965}, therefore, the equations in the following are specialized to a crystal with two non equivalent axis, the $a$ (along the $x$ or $y$ direction) and $c$ (along the $z$ direction) axis. For a uniaxial strain along the $a$ or $c$-axis the strain tensor~is, 
\begin{equation}
\uuline{\epsilon}^{(a)}  = \begin{pmatrix} \epsilon_{xx} & 0 & 0 \\ 0 & 0 & 0 \\ 0 & 0 & 0 \end{pmatrix}  \:;\qquad \uuline{\epsilon}^{(c)}  = \begin{pmatrix} 0 & 0 & 0 \\ 0 & 0 & 0 \\ 0 & 0 & \epsilon_{zz} \end{pmatrix}\:,
\end{equation}
where typically ${\epsilon_{xx}\ll 1}$ and ${\epsilon_{zz} \ll 1}$. Under these conditions, a linear elastic behavior can be assumed and the terms $\mathcal{O}(\epsilon^3)$ in Eq.\:(\ref{eq:UoverV}) can be neglected. Furthermore, due to the fact that all linear thermal-expansion coefficients shown in Fig.\:\ref{fig:TECs_SiO2_CaWO4} are very small at temperatures $\leq 10$\:K, we can approximate the amplitude of the linear strain as,
\begin{equation}
\label{eq:ltec}
\epsilon_{\mu,T_2^\pI} = \frac{L_{\mu}^\pI(T_2^\pI)-L_{\mu}^\pI(T_1^\pI)}{L_{\mu}^\pI(T_1^\pI)}\:,
\end{equation} 
where $\mu=xx$ ($zz$) for strain along the $a$ ($c$)-axis, $L_{\mu}^\pI(T)$~is~the length of the absorber in $\mu$ direction at temperature $T$, and {$T_1^\pI \leq 10$\:K}.
Assuming that we know $L_{\mu}^\pI(T_2^\pI)$, we can compute  $L_{\mu}^\pI(T_1^\pI)$ using Eq.\:(\ref{eq:alpha}) as,
\begin{equation}
L_{\mu}^\pI(T_1^\pI) = L_{\mu}^\pI(T_2^\pI) \exp\!\left[{-{\int_{T_1^\pI}^{T_2^\pI}\!\!\alpha (T)\, \d T}}\right]\:,
\label{eq:integral}
\end{equation}
Being solely focused on the elastic-energy change brought on by a temperature shift from $T_1^\pI$\:=~10\:K to $T_2^\pI$\:=~60\:K, we can assume the reference elastic energy $U_{\mu}^\pI(T_1^\pI)$ to be zero, while the energy $U_{\mu}^\pI(T_2^\pI)$ can be given as
\begin{equation}
U_{\mu}^\pI(T_2^\pI) = u\!\left(\epsilon_{\mu,T_2^\pI\!}\right) V(T_2^\pI)\:,
\end{equation}
with 
\begin{equation}
\label{eq:volume}
V(T)=\left[L_{xx}^\pI(T)\right]^{\hspace{-0.1mm}2} L_{zz}^\pI(T)\simeq V(T_1^\pI)\:,
\end{equation}
being the volume of the absorber at temperature $T$. The last approximation in Eq.\:(\ref{eq:volume}) is well justified because both
$\max[L_\mu^\pI(T)-L_\mu^\pI(0\:K)] \sim \mathcal{O}(10\:\upmu\mbox{m})$ for CaWO$_4$ at all temperatures of interest (see Fig.~\ref{fig:cooldown}, third panel)
and all linear dimensions of the absorber are around 1\:cm, so our setup (according to Eq.\:(\ref{eq:ltec})) satisfies the condition of very small strain. Using Eqs.\:(\ref{eq:ltec}) to (\ref{eq:volume}), the variation of the elastic energy, defined as ${\Delta U_{\mu}^\pI \equiv U_{\mu}^\pI(T_2^\pI) - U_{\mu}^\pI(T_1^\pI)}$, can be expressed~as,
\begin{equation}
\label{eq:U}
\Delta U_{\mu}^\pI 
 \simeq \frac{1}{2} \, C_{\mu\mu}^\pI \left[{L}_{\mu}^\pI(T_2^\pI)-{L}_{\mu}^\pI(T_1^\pI)\right]^{\!2} {L}_{\mu}^\pI(T_1^\pI)\:.
\end{equation}

Using the results for CaWO$_4$ from~\cite{senyshyn2004lattice}, we integrate, as in Eq.\:(\ref{eq:integral}), the termal-expansion coefficients after fitting the data with a fourth-degree polynomial function. The elastic-stiffness constant corresponding to strain along the $a$-axis ($c$-axis) is $C_{xx,xx}$ ($C_{zz,zz}$), or $C_{11}$ ($C_{33}$) in Voigt notation. The values of these elastic constants for CaWO$_4$ at 77\:K are $C_{11}$\:=\:149\:GPa and $C_{33}$\:=\:132\:GPa~\cite{FARLEY1971965,Gluyas_1973}, with a small increase at lower temperatures, namely $C_{11}$\:=\:151\:GPa and $C_{33}$\:=\:132\:GPa at 4.2\:K~\cite{Gluyas_1973}. Including the values of the elastic constant at low temperatures in Eq.\:(\ref{eq:U}) and calculating $L_{\mu}^\pI$ using the TEC for the $a$ and $c$ axis, we find that a cooldown from 60 to 10\:K for CaWO$_4$ corresponds to an energy release\footnote{The variation of energy in the cooldown would be negative. A negative energy variation corresponds to a release of energy.} equal to about 2.8\:MeV (0.3\:MeV) for the $c$-axis ($a$-axis) orientation.~Integrating the energy spectrum of a CRESST-III module of CaWO$_4$ (TUM93A) measured after a 60\:K warm-up of the cryostat results\footnote{CRESST-III, private communication based on~\cite{dissertation}} in a value of about 20\:MeV. The order of magnitude of the computed total elastic energy associated to the strain compared to the total energy of the LEE events after a warm-up shows that elastic stress processes could play a role in the generation of the LEE events. This simplified calculation motivates further investigations of this hypothesis.

Inelastic terms and a microscopic description of the interface must be introduced to describe stable displacements at the interface induced by the relative TEC mismatch, namely \textit{dislocations}. A microscopic treatment of the dislocation at the atomic level is required to translate the macroscopic description into effects involving phonon generation. To this purpose, simulations based on molecular dynamics and density-functional theory are ongoing. 



\section{Interfacial dislocation nucleation}
\label{sec:dislocation}
It is known that relative thermal contraction of materials with different expansion coefficients can induce dislocation nucleation~\cite{ARSENAULT1986175}. Dislocation nucleation is largely investigated in literature in the context of epitaxially grown thin films, see, \eg,~\cite{OHRING2002417}. A study of surface dislocations~\cite{PhysRevLett.100.025502} reports on the activation free energy of surface dislocation nucleation of a crystalline solid under constant temperature and strain rate and defines the crystal surface itself as a defect with respect to the perfect crystal. Dislocation generation induced by thermal-expansion mismatch (called misfit in material science) is shown to occur also in the bulk of crystals around \textit{inclusions}~\cite{ARSENAULT1986175}\footnote{Quoting~\cite{ARSENAULT1986175}: \textit{``[$\dots$] when the composite is cooled from the elevated temperatures of annealing or processing, misfit strains which are sufficient to generate dislocations occur because of differential thermal contraction at the Al-SiC interface.''}}.\\[-3mm]

\paragraph{Nucleation rate} The rate of surface dislocation nucleation, $\nu$, is given as~\cite{PhysRevLett.100.025502},
\begin{equation}
\nu = N \, \nu_0\, \exp\!\left[-\frac{Q(\sigma, T)}{k_B T}\right],
\end{equation}
where $N$ is the number of equivalent nucleation sites,  $\nu_0$~is~the \textit{attempt frequency} (or average phonon frequency), $k_BT$ is the thermal energy, and $Q$ is the activation barrier energy, depending on stress $\sigma$ and temperature. Using atomistic simulations (\eg, based on molecular dynamics or density-functional theory), the activation energy can be estimated and by measuring the number of initial defects using for example atomic force microscopy (AFM) or transmission electron microscopy (TEM), the rate of dislocation as a function of temperature could be estimated.\\[-3mm]

\paragraph{Dislocation topology} In~\cite{roos1994thermal} the relation between thermal-expansion mismatch and thermal stress is discussed and using a germanium/silicon interface it is found that stress-induced dislocations nucleate as half-loops at the surface. The results show that nucleation is more likely when other defects or contamination are already present on the surface~\cite{roos1994thermal}. In our case of study, first surface dislocation nucleation could be induced by the thin film fabrication process~\cite{armour1992surface}, while further dislocation could nucleate during the cooldown due to the thermal-expansion mismatch of the interfaced materials.\\[-3mm]

\paragraph{Threshold dislocation energy (TDE)}
In~\cite{ijms24043289} the effect of strain on the TDE for a material's radiation damage is shown to depend on the strain that the material undergoes and on the crystal orientation. The TDE for their case of study which was tantalum and tantalum-tungsten alloy, is $\mathcal{O}$(10)\:eV.

These studies in the field of material science and engineering support the hypothesis that the thermal-expansion coefficient mismatch at interfaces causes the formation of dislocations right at the interface of the cryogenic calorimeters targets and the sensor thin-film fabrication layers. The work in~\cite{ARSENAULT1986175} on the thermal-expansion mismatch in the bulk around inclusions could also play a role in the explanation of the component of the LEE scaling with mass~\cite{TESSERACT:2025odn}.

\section{Interfacial thermal-boundary resistance effect in double-TES modules}
\label{sec:acma}

Recently, Raya-Moreno and co-workers~\cite{Raya-Moreno:2023fiw} highlighted that phonon transmission between interfaces is ruled by the matching of the phonon dispersion curves of the interfaced materials. A phonon with frequency higher than the available phonon band in the phonon collector (PC) would be reflected at the interface. This effect occurs for example in a Al$_2$O$_3$/Al interface, where the highest modes of Al$_2$O$_3$ are at about 30\:THz and the highest modes of Al at about 10{\thinspace}THz (see~\cite{Raya-Moreno:2023fiw}, Figure~5). This phenomenology has a clear implication on the surface rejection power of double-TES modules.

\begin{table}
\label{tab:frequencies}
\begin{center}
\caption{List of approximated values of the highest phonon-band energies for crystals and phonon collectors (PC). $1\thinspace\mbox{THz} \simeq  4.13\thinspace\mbox{meV}$.}
\vspace{1mm}
\begin{tabular}{ l  c c c| c c c c}
\hline
 Crystal\;\;\; & \;THz\; & \;\;meV\;\; & \;Ref.\; & 
\;\;PC\;       & \;THz\; & \;\;meV\;\; & \;Ref.\; \\ \hline
 Al$_2$O$_3$ & 27 & 112 & \cite{Raya-Moreno:2023fiw} & Al & 10 & 41 & 
 \cite{Raya-Moreno:2023fiw}\\  
 CaWO$_4$    & 27 & 112 & \cite{senyshyn2004lattice} & Cu & 8  & 33 & \cite{smirnov2020copper}\\
 Si          & 15 & 62  & \cite{jain2013commentary}  & Au & 5  & 21 & \cite{smirnov2020copper}\\ 
 GaAs        & 8  & 33  & \cite{jain2013commentary}  & W  & 7  & 29 & \cite{jani1976phonon}\\
 NaI         & 5  & 21  & \cite{jain2013commentary}  &    &\\
\hline
\end{tabular}
\end{center}
\end{table}

In the context of the double-TES modules which have been proposed to perform surface background rejection~\cite{ GRAPES-3:2024yym, TESSERACT:2025odn}, if a relaxation event from the shear stress (or other sources) occurs at the interface between absorber and TES, the high-energy optical or acoustical phonons are reflected by the closest TES, down-convert in the absorber and propagate through the target until they are absorbed into a matching available state in the phonon collector. Due to the propagation, such events could be detected as shared events in the two TESs. This would imply that LEE events generated in the absorber by thermal-expansion mismatch between sensor and absorber cannot be rejected using double-TES modules if the sensor and absorber have a mismatch in the phonon-dispersion curves with higher modes in the absorber. We report a list of crystals and phonon collectors associated to their highest phonon frequency in Table~I.

\section{Possible tests and solutions}
\label{sec:proposal}
We propose three detector module designs to address the effect of both the interfacial thermal-expansion mismatch and the interfacial frequency mismatch on the LEE, which modify the current double-TES designs.

\begin{figure}[h]
\vspace{-4mm}
\includegraphics[width=\linewidth]{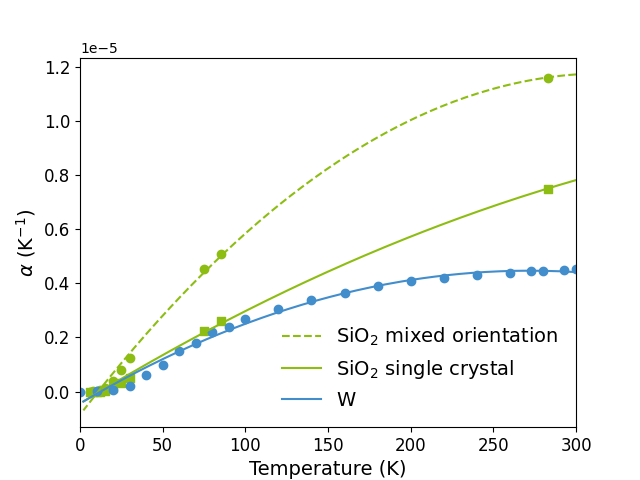}
\vspace{-6mm}
\caption{Thermal-expansion coefficients of SiO$_2$ and W as a function of temperature.} 
\label{fig:TECs_SiO2_W}
\end{figure}

\begin{figure}
\vspace{-4mm}
\includegraphics[width=\linewidth]{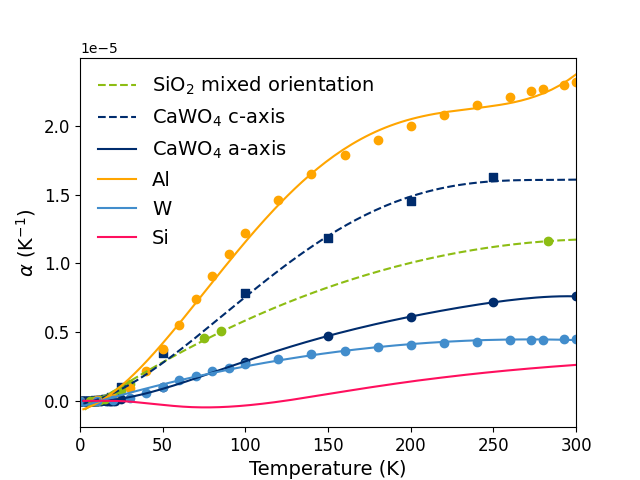}
\vspace{-6mm}
\caption{Thermal-expansion coefficients of SiO$_2$, CaWO$_4$, Al, W, and Si as a function of temperature.} 
\label{fig:TECs_SiO2_W_CaWO4}
\end{figure}
 
\subsection{Double $\alpha$-SiO$_{2,\parallel}$/W-TES}
SiO$_2$ has a crystalline form which can be grown along a perpendicular, parallel or mixed axes with respect to the planar basis~\cite{white1964thermal}. The thermal-expansion coefficient of these three forms and the one of W are available in literature, we report them in Fig.~\ref{fig:TECs_SiO2_W}. The comparison shows that the crystalline form of SiO$_2$ oriented along the parallel axes has a thermal expansion coefficient similar to the one of W below 100\:K. We propose to fabricate a double-TES on SiO$_{2,\parallel}$, made of only W, with no amorphous SiO$_2$ in between and perform a slow cooldown. R\&D to fabricate the W film on the target surface which minimizes the temperature jumps is also necessary, to avoid the formation of the first lattice dislocations.

\subsection{Double CaWO$_{4,a}$/W-TES}
The $a$-crystallographic axis orientation of CaWO$_4$ has a thermal expansion similar to the one of W, we compare them in Fig.~\ref{fig:TECs_SiO2_W_CaWO4}. We propose to use a CaWO$_4$, oriented along the $a$-axis and equipped with a W-only (without Al phonon collectors or any other film on the surface) double-TES. If the thermal-expansion coefficient mismatch contributed to the LEE, a slow cooldown until 100\:K and the use of this module will result in a reduced~LEE.

\subsection{Double GaAs-TES}
The highest phonon-dispersion curve of GaAs corresponds to a frequency of about 8\:THz (see Table~I
and Fig.\:2 of~\cite{giannozzi1991ab}). This module would be still affected by the thermal-expansion coefficient mismatch. However, the better acoustic matching of GaAs with the possible phonon collectors, if no additional amorphous layer is added in between, is expected to result into a ``migration" of events from the shared event band to the single event bands of a double-TES and thus in a higher rejection power for surface events. 

\section{Conclusions}
The low-energy excess observed by CRESST-III could originate from the thermal-expansion mismatch at the interface between the crystal absorber and the SiO$_2$ amorphous layer underneath the sensor. In this work we use a simple model to show that the thermoelastic energy released during a cooldown of a crystal absorber is of comparable order of magnitude to the total energy of LEE events observed in a CRESST-III warm-up test. We show that the discrimination power of the double-TES modules built to reject surface events may suffer from the frequency mismatch between absorbers and sensors, since high-energy phonons generated at the interface would reflect at the interface and propagate through the crystal and thus reach the second TES. We propose three double-TES modules which would mitigate or solve these problems, namely a $\alpha$-SiO$_{2,\parallel}$ double-TES made from only tungsten, a CaWO$_{4}$ crystal oriented along the $a$-axis, also equipped with W-only double-TESs, and a standard GaAs double-TES which has a better frequency matching and, thus, an expected improved phonon-background discrimination power. 

\begin{acknowledgments}
This work would have not been possible without the detailed work performed by the CRESST-III collaboration. We thank Dominik Fuchs and Jochen Schieck for reviewing the manuscript and providing useful comments, and  Anatoliy Senyshyn for sharing additional data points for the simulated TEC of CaWO$_{4}$ at low temperature. We are thankful to the Klaus Tschira Foundation for supporting this research. 

\end{acknowledgments}

\begin{appendix}

\end{appendix}

\bibliography{refs.bib}

\end{document}